# Giant spin-torque diode sensitivity at low input power in the absence of bias magnetic field

Bin Fang[1], Mario Carpentieri[2], Xiaojie Hao[3], Hongwen Jiang[3], Jordan A. Katine[4], Ilya N. Krivorotov[5], Berthold Ocker[6], Juergen Langer[6], Kang L. Wang[7], Baoshun Zhang[1], Bruno Azzerboni[8], Pedram Khalili Amiri[7*], Giovanni Finocchio[8*] and Zhongming Zeng[1*]

[1]Key Laboratory of Nanodevices and Applications, Suzhou Institute of Nano-tech and Nano-bionics, Chinese Academy of Sciences, Ruoshui Road 398, Suzhou 215123, P. R. China

[2]Department of Electrical and Information Engineering, Polytechnic of Bari, Bari 70125, Italy

[3]Department of Physics and Astronomy, University of California, Los Angeles, California 90095, United States

[4]HGST Inc., 3403 Yerba Buena Road, San Jose, California 95135, United States

[5]Department of Physics and Astronomy, University of California, Irvine, California 92697, United States

[6]Singulus Technologies, Kahl am Main, 63796, Germany

[7]Department of Electrical Engineering, University of California, Los Angeles, California 90095, United States

[8]Department of Electronic Engineering, Industrial Chemistry and Engineering, University of Messina, Messina 98166, Italy

*email: gfinocchio@unime.it, pedramk@ucla.edu or zmzeng2012@sinano.ac.cn

**Microwave detectors based on the spin-transfer torque diode effect are among the key emerging spintronic devices. By utilizing the spin of electrons in addition to charge, they have the potential to overcome the theoretical performance limits of their semiconductor (Schottky) counterparts, which cannot operate at low input power. Here, we demonstrate nanoscale magnetic tunnel junction microwave detectors, exhibiting record-high detection sensitivity of 75400 mVmW$^{-1}$ at room temperature, without any external bias fields, for input microwave power down to 10 nW. This sensitivity is 20× and 6× larger than the state-of-the-art Schottky diode detectors (3800 mVmW$^{-1}$) and existing spintronic diodes with greater than 1000 Oe magnetic bias (12000 mVmW$^{-1}$), respectively. Micromagnetic simulations supported by microwave emission measurements reveal the essential role of the injection locking mechanism to achieve this sensitivity performance. The results enable dramatic improvements in the design of low-input-power microwave detectors, with wide-ranging applications in telecommunications, radars, and smart networks.**

I. Introduction

The continuous progress in the development of magnetic materials and nanostructures has enabled spintronic devices with performance superior to semiconductor-based electronics, offering promising solutions for a range of future high-speed and energy-saving electronic systems [1, 2, 3, 4, 5, 6, 7]. In particular, the spin-transfer torque (STT)[8, 9] induced by a d.c. spin-polarized current can switch the magnetization[10, 11], or excite self-oscillations[12, 13], in magnetic nanodevices, giving rise to applications such as memories and nanoscale oscillators [14, 15, 16]. On the other hand, microwave detectors (rectifiers) based on the STT-diode effect can be realized when the d.c. input is replaced by a microwave current[17]. The STT-diode effect is the result of STT-induced ferromagnetic resonance, which leads to a rectification effect (i.e. d.c. voltage) in magneto-resistive nanoscale devices. Since its discovery[17], this effect has been used for quantitative measurements of magnetic torques in magnetic tunnel junctions (MTJs) (e.g. Slonczewski, field-like, and voltage-controlled torques) [18, 19, 20, 21, 22].

For practical applications in microwave detectors, the ability to obtain a device with high detection sensitivity without using external magnetic fields, at room temperature and at low input microwave powers is crucial. However, there are currently no practical (spintronic or conventional semiconductor) solutions that achieve all of these requirements simultaneously. In STT diodes studied to date, the application of an additional external magnetic field (often canted at an angle with respect to the device plane) is required to achieve large microwave detection sensitivity[21, 23, 24, 25, 26, 27, 28]. Although this external field may be integrated into the device in principle, e.g. by engineering the material stack, or by using coils or permanent magnets, it is undesirable from a practical point of view due to increased size and cost. On the other hand, semiconductor-based Schottky diode detectors, while not requiring a magnetic bias, fail to offer sufficient sensitivity for low input microwave powers, principally due to their high zero-bias junction resistance[29].

In this work, we present a STT-diode microwave detector meeting all of the above-mentioned requirements, with an ultrahigh detection sensitivity of 75400 mVmW$^{-1}$. This is achieved by incorporating three elements into the device. First, a perpendicularly magnetized free layer[6, 7, 30], which allows for device operation in the absence of external magnetic fields[31]. Secondly, the injection locking mechanism [32] due to the simultaneous application of a d.c. bias current and the input microwave signal, which enhances the detection sensitivity. Thirdly, an MgO-based MTJ material stack exhibiting high tunnel magnetoresistance (TMR). The operation mechanism and the fundamental role of the injection locking are discussed based on microwave emission measurements combined with micromagnetic simulations. Importantly, unlike Schottky diodes, the ultrahigh sensitivity is observed even at a very low input microwave power of 10 nW. The measured detection sensitivity is about 20× larger than the one in state-of-the-art Schottky diode detectors (3800 mVmW$^{-1}$)[33], and more than 6× better than existing spintronic diodes (12000 mVmW$^{-1}$), which additionally require large magnetic bias fields for their operation[28].

## II. Experiments and Discussion

The devices studied in this work have an MTJ structure consisting of a synthetic antiferromagnetic Co$_{70}$Fe$_{30}$ (2.3 nm)/Ru (0.85 nm)/Co$_{40}$Fe$_{40}$B$_{20}$ (2.4 nm) reference layer, exchange biased by a PtMn film, designed to have an in-plane easy axis, and a Co$_{20}$Fe$_{60}$B$_{20}$ perpendicularly magnetized free layer, separated from the reference layer by a 0.8 nm MgO tunnel barrier. A schematic of the device is shown in Fig. 1a. The free layer has an out-of-plane easy axis at zero bias field, which is realized by tuning the perpendicular magnetic anisotropy at the interface of the Co$_{20}$Fe$_{60}$B$_{20}$ layer with the MgO tunnel barrier[6, 7, 30, 31]. This magnetization configuration enables the device to excite large-amplitude magnetization precession under a small STT[31]. In addition, the Co$_{20}$Fe$_{60}$B$_{20}$-MgO-Co$_{40}$Fe$_{40}$B$_{20}$ material combination ensures high TMR[4, 5, 6, 7]. These factors are the key ingredients to enhance the sensitivity of the STT-diode. Electron-beam lithography and ion milling were used to fabricate the pillar-shaped devices with 150 nm × 60 nm elliptical cross-section. Qualitatively similar results were also obtained in devices with other dimensions (130 nm × 50 nm and 150 nm × 70 nm) as well.

Figure 1b shows an example of the resistance dependence as a function of the in-plane magnetic field applied parallel to the ellipse major axis ($H_x$). As the field increases from -1000 Oe to +1000 Oe, the resistance increases gradually as the free-layer magnetization is aligning antiparallel to the reference layer magnetization. At zero field, a small tilting angle ($\theta = 76°$) from the out-of-plane configuration is measured due to the coupling between the free and reference layers. We estimated $\theta$, the angle between the magnetization vectors of the free layer and the reference layer, from the MTJ resistance[28]

$$R^{-1}(\theta) = \frac{R_P^{-1} + R_{AP}^{-1}}{2} + \frac{R_P^{-1} - R_{AP}^{-1}}{2}\cos(\theta) , \qquad (1)$$

where the resistances in anti-parallel ($R_{AP}$) and parallel ($R_P$) configurations are 1200 $\Omega$ and 640 $\Omega$, respectively. In addition, the free layer exhibits a voltage-controlled interfacial perpendicular magnetic anisotropy (VCMA) estimated to be 34 fJV$^{-1}$m$^{-1}$ (see Supplementary Materials, section S1), in good agreement with previous reports for similar material structures[21, 22, 34, 35]. All measurements reported below were carried out at room temperature and zero bias magnetic field.

We studied the STT diode response by using ferromagnetic resonance (FMR) measurements as shown in Fig. 1a[18]. In this technique, a weak microwave current $I_{ac}\sin(2\pi f_{ac}t)$ and a d.c. current $I_{dc}$ were applied to the device using a bias Tee. When a microwave current at a frequency $f_{ac}$ is applied, the free layer magnetization starts to precess, resulting in a time-dependent resistance oscillation due to the TMR effect. As a result, a rectified voltage is generated across the MTJ[17, 18]. To improve signal-to-noise ratio, the microwave input was modulated at a low frequency (10 kHz) and the resulting rectified voltage was measured with a lock-in amplifier. Figure 2a shows the measured rectified voltage as a function of the microwave frequency at $I_{dc} = 0$ mA for input microwave powers ranging from 3.2 to 100 nW. The maximum voltage is measured at the resonant frequency $f_0 = 1.2$ GHz. The FMR spectra in Fig. 2a are well fitted by a sum of symmetric and antisymmetric Lorentzian functions with identical resonant frequency $f_0$. The origin of the asymmetric lineshape is related to the VCMA effect[21]. A bias-field-free detection sensitivity, defined as the rectified voltage divided by the input microwave power i.e. $V_{detect}/P_{RF}$, was obtained to be 970 mVmW$^{-1}$, which is larger than previously reported values (440 mVmW$^{-1}$ and 630 mVmW$^{-1}$)[21, 28] for MTJ-based STT diode detectors, and at a comparable level to unbiased commercial Schottky diode detectors (500-1000 mVmW$^{-1}$)[33].

We next conducted the STT diode response studies under different d.c. bias currents. Figures 2b and 2c show the detected voltage curves as a function of the microwave frequency, at a low input microwave power of 10 nW, for the range of d.c. bias current from -0.40 to +0.25 mA. Positive d.c. currents were found to suppress the detection voltage, because STT increases the damping of the magnetization precession and the perpendicular anisotropy in the free-layer. For a range of negative currents (-0.32 < $I_{dc}$ < -0.22 mA), the detection voltages were significantly enhanced, while for large negative currents ($I_{dc}$ < -0.35 mA) no well-defined detection spectra were observed, which may relate to a destabilization of the anti-parallel state by the STT[28]. Figure 2d summarizes the maximum rectified voltage as a function of the bias current. At $I_{dc}$ = -0.25 mA the maximum detected voltage reaches 754 µV, which is about 80× larger than the one measured at zero bias current. The corresponding detection sensitivity is 75400 mVmW$^{-1}$, nearly 20× larger than the state-of-the-art biased Schottky diode detectors (3800 mVmW$^{-1}$)[33], and substantially larger than the best existing spintronic diode (12000 mVmW$^{-1}$)[28]. Figure 3 compares our results with previously reported rectified voltage sensitivities for different STT diodes, with the indication of the required external magnetic field. Here, the preferred (top-left) corner represents the combination of high detection sensitivity and bias-field-free operation. The inset of Fig. 2d displays the sensitivity as a function of the input microwave power at different values of current ($I_{dc}$ = 0, -0.10, -0.225, -0.25 and -0.275 mA). It is worth noting the fact that the observed giant detection sensitivities are not only substantially larger than those of Schottky diodes, but they are achieved at very low microwave input power (< 100 nW, and down to a few nW). By comparison, Schottky diodes typically do not provide satisfying microwave-to-dc conversion efficiency for sub-µW input microwave power, mainly due to their high zero-bias junction resistance[29]. In addition, the STT diode devices can be scaled down to nano-meter size (0.07 μm$^2$ in this study), which makes it potentially suitable for compact on-chip microwave detectors. Hence our results demonstrate a pathway to design competitive MTJ-based STT diodes for the replacement of Schottky diodes in a wide range of applications, such as energy efficient telecommunication devices, radars, and smart networks.

We next focus on the mechanisms enabling the giant detection sensitivity. The present results cannot be explained in terms of the theory developed for microwave detectors working in the out-of-plane regime[36], in which the device operates as a non-resonant broadband microwave detector, and the output voltage is virtually independent of the input microwave power. The device discussed here exhibits a resonant character (see Figs. 2b and 2c), and the output power depends on $P_{RF}$ (see inset of Fig. 2d). An alternative mechanism is the non-linear FMR[28], where both negative and positive d.c. bias currents have shown to enhance the detection voltage while not changing the detection frequency. By contrast, the device presented here exhibits the enhanced detection voltage only at negative currents, along with a tunability of the resonant frequency on current (see Fig. 2b, 2c or Fig. 5b). Hence, the underlying physics in this study appears to be different from the ones of previous works.

## III. Micromagnetic Simulations and the Effect of Injection Locking on the Detection Sensitivity

To gain a deeper understanding of the ultrahigh sensitivity mechanism, we performed micromagnetic simulations (see Methods). Importantly, our computations show persistent magnetization dynamics (self-oscillations) (at $I_{ac}$ = 0 mA) and injection locking[32] (at $I_{ac}$ > 0 mA) for the same range of d.c. currents. Figure 4a shows a comparison between the self-oscillation frequency $f_p$ (olive color) and the frequency $f_0$ (orange color) of the maximum rectification voltage, as a function of the d.c. bias current. The self-oscillation frequency exhibits a red shift with increasing negative current, with a frequency jump of about 300 MHz near the current value of $I_{dc}$ = -0.25 mA. In the locking region, the d.c. bias current excites a persistent precession of the free-layer magnetization, which is locked to the input microwave frequency. Figure 4b displays an example of the microwave emission frequency $f_p$ as a function of the frequency of the input microwave current, for $I_{dc}$ = -0.26 mA and $I_{ac}$ = 0.07 mA. In the locking region between 510 and 540 MHz, the time-dependent resistance is given by $R(t) = R_0 + \Delta R_s \sin(2\pi f_{ac} t + \Phi_s)$, $\Phi_s$ and $\Delta R_s$ are the intrinsic phase shift between the microwave current and the oscillating TMR signal[37] and the amplitude of the TMR, respectively. The corresponding rectified voltage $V_{detect} = \frac{1}{2} \Delta R_s I_{ac} \cos(\Phi_s)$ is shown in Fig. 4c. We stress the fact that the value of $\Delta R_s$ can be significantly larger than the case with no bias current, given that it results from the amplitude of the self-oscillation of the magnetization, rather than conventional ferromagnetic resonance, and its value is almost independent of $f_{ac}$ inside the locking region. Consequently, the larger value of the maximum detection voltage, which occurs at $\Phi_s = 0$, is due to the large value of $\Delta R_s$. The origin of $\Phi_s$ can be understood from the theory developed in Ref. 38, in which it is directly linked to the coupling between the amplitude and the frequency of the self-oscillation (see Eq. 56 in Ref. 38). The inset of Fig. 4b shows the computed locking region as a function of $I_{ac}$. The injection locking results are qualitatively similar to the ones presented in Refs. 39, 40, in which a microwave current is used to optimize the microwave emission properties induced by a d.c. current, such as linewidth and output power. It is noted that the value of $\Phi_S$ inside the locking bandwidth depends on the microwave frequency (Fig. 4d), similar to that in a previous study[41]. This analysis points to a novel scenario where the injection locking achieved at low microwave currents is responsible for the observed giant detection sensitivity.

To further verify the role of injection locking, we conducted additional experiments to study the current-induced microwave emission[12] in our samples, recorded using a 9 kHz-26.5 GHz spectrum analyzer (see Supplementary Materials, section S2). Figure 5a summarizes the microwave emission data as a function of the d.c. current. The oscillation frequency exhibits a red shift with increasing bias current. At $I_{dc}$ = -0.25 mA, a 300 MHz jump in the oscillation frequency is measured, along with a reduction of the output microwave power. The d.c. resistance measurement (Fig. 5c) shows that the frequency jump is related to the change of sign of the *x*-component of the oscillation axis from positive to negative, with a consequent change in the symmetry of the STT. Similar results were also observed at low out-of-plane bias fields[42]. Near $I_{dc}$ = -0.4 mA the steady magnetization dynamics is switched off, analogous to the disappearance of the detection spectrum in the diode measurements (see Fig. 2c). Figure 5b shows a comparison between the precession frequency $f_p$ of the current-induced microwave emission and the frequency $f_0$ of the maximum rectification voltage as a function of the bias current. Two different regions can be identified: (i) $I_{dc}$ > -0.21 mA and -0.4 < $I_{dc}$ < -0.25 mA, where the two frequencies are very close, and (ii) -0.25 < $I_{dc}$ <-0.21 mA, where 300 MHz difference in two frequencies (same as the frequency jump achieved at $I_{ac}$ = 0 mA) is observed. Those results are quantitatively consistent with the simulation data (compare Fig. 5b and Fig. 4a) in region (i). On the other hand, the existence and current range of the region (ii) exhibits a slight device to device variation (not shown). The difference in the two frequencies $f_p$ and $f_0$ in region (i) can be explained as follows. When the microwave frequency is equal to the precession frequency ($f_{ac} = f_p$), the phase shift $\Phi_s$ of the detection voltage is different from zero[37, 38] as also seen in Fig. 4d. As a result, the frequency of the maximum detection voltage and the precession frequency for a fixed $I_{dc}$ do not coincide. This comparison confirms the hypothesis that injection locking governs the behavior of the ultrasensitive microwave detector. In other words, for a specific frequency range, i.e. the locking bandwidth, the frequency of the microwave emission is locked to the microwave input frequency, thereby enhancing the detection sensitivity.

In summary, we have demonstrated giant sensitivity of nanoscale spintronic diodes in the absence of any external magnetic field, at room-temperature and for low input microwave power. The achieved detection sensitivity of 75400 mV mW$^{-1}$ is 20× higher than that of semiconductor Schottky diodes. The analysis of micromagnetic simulations and the microwave emission measurements reveal that the injection locking is the key mechanism to achieve our results. It is expected that these findings will move spintronic microwave detectors considerably closer to industrial application.

**Methods**

**Sample preparation**. The continuous multilayer thin films with stacks of composition PtMn (15) /Co$_{70}$Fe$_{30}$ (2.3)/Ru (0.85)/Co$_{40}$Fe$_{40}$B$_{20}$ (2.4)/MgO (0.8)/Co$_{20}$Fe$_{60}$B$_{20}$ (1.63) (thickness in nm) were deposited using a Singulus TIMARIS physical vapor deposition system and annealed at 300°C for two hours in a magnetic field of 1 T. The films were subsequently patterned into ellipse-shaped pillars using electron-beam lithography and ion milling techniques. The resistance-area product in the parallel magnetization configuration was 4.5 Ω•μm$^2$, and the in-plane TMR ratio, defined as ($R_{AP}$-$R_P$)/$R_P$, was 87.5%.

**Micromagnetic simulations.** We numerically solve the Landau-Lifshitz-Gilbert-Slonczewski equation which includes the field-like torque $T_{OP}$[43, 44], and the voltage dependence of the VCMA[21]. The $T_{OP}$ is considered to be dependent on the square of the bias voltage up to a maximum value of 25% of the in-plane torque computed for a current density $|J|$ = 4.0×10$^6$A/cm$^2$.[19] The total torque, including also the in-plane component $T_{IP}$ is given by

$$T_{IP} + T_{OP} = \frac{g}{|e|\gamma_0} \frac{|\mu_B| J(\mathbf{m},\mathbf{m_p})}{M_s^2 t} g_T(\mathbf{m},\mathbf{m_p}) \left[ \mathbf{m} \times (\mathbf{m} \times \mathbf{m_p}) - q(V)(\mathbf{m} \times \mathbf{m_p}) \right], \tag{2}$$

where $g$ is the gyromagnetic splitting factor, $\gamma_0$ is the gyromagnetic ratio, $\mu_B$ is the Bohr magneton, $q(V)$ is the voltage-dependent parameter for the perpendicular torque, $J(\mathbf{m},\mathbf{m_p})$ is the spatially nonuniform current density, $V$ is the voltage (computed from $I_{dc}$-$R$ curves, see Fig. 5c), $t$ is the thickness of the free layer, and $e$ is the electron charge. The effective field takes into account the standard micromagnetic contributions (exchange, self-magnetostatic) as well as the Oersted field due to both microwave and d.c. current. The presence of the VCMA has been modeled as an additional contribution to the effective field. The parameters used for the CoFeB are saturation magnetization $M_s = 9.5 \times 10^5$ Am$^{-1}$, exchange constant $A = 2.0 \times 10^{-11}$ Jm$^{-1}$, and damping parameter $\alpha = 0.02$.[31] The zero bias VCMA constant $k_u = 5.52 \times 10^5$ Jm$^{-2}$ has been estimated from the fitting of the FMR frequency $f_0 = 1.245$ GHz from Fig. 2a, while the VCMA coefficient is 34 fJV$^{-1}$m$^{-1}$ (Supplementary Materials, section S1). The minimum value of the $k_u$ achieved at $I_{dc} = -0.35$ mA is $5.45 \times 10^5$ J m$^{-2}$. The polarization function $g_T(\mathbf{m},\mathbf{m_p}) = 2\eta_T(1+\eta_T^2 \mathbf{m} \cdot \mathbf{m_p})^{-1}$, where $\mathbf{m}$ and $\mathbf{m_p}$ are the normalized magnetizations of the pinned and free layer, has been computed by Slonczewski[45, 46]. We have used for the spin-polarization $\eta_T$ the value 0.6.[19]

**Acknowledgements**

This work was supported by the 100 Talents Programme of The Chinese Academy of Sciences and the National Science Foundation of China (11274343, 11474311), the DARPA STT-RAM and Non-Volatile Logic programs, and the Nanoelectronics Research Initiative (NRI) through the Western Institute of Nanoelectronics (WIN). The work at UCLA was also supported in part by the NSF Nanosystems Engineering Research Centre for Translational Applications of Nanoscale Multiferroic Systems (TANMS). This work was also supported by the project PRIN2010ECA8P3 from Italian MIUR and NSF support through grants DMR-1210850, DMR-1124601 and ECCS-1309416.

**Author contributions**

B. F., Z. M. Z., G. F. and H. W. J. initiated this work. P. K. A. and I. N. K. designed the MTJ devices. J. L. and B. O. deposited the films, and J. A. K. fabricated the devices. B. F. and X. J. H. performed the measurements, M. C. and G. F. performed micromagnetic simulations. Z. M. Z. and G. F. analysed the data and wrote the paper with contributions from P. K. A. All authors contributed to the discussion and commented on the manuscript.

**Competing financial interests**

The authors declare no competing financial interests.

**Figure captions**

**Figure 1 | Spin-torque diode device**. **a**, Spin-torque diode device and schematic of circuit used for FMR measurements. The device is based on an MTJ with an in-plane magnetized reference layer and a perpendicularly magnetized free layer separated by a MgO tunnel barrier. A microwave input from a signal generator (E8257D, Agilent Technologies) and a d.c. bias from source meter (2400, Keithley) were applied through a bias tee. The voltage detection ($V_{detect}$) is measured by a low-frequency (10 kHz) modulation method using a lock-in amplifier (SR830, Standard Research Systems). **b**, The magnetoresistance curve of the MTJ device under in-plane magnetic fields for $I_{dc}$ = 10 µA.

**Figure 2 | Voltage detection characteristics of STT diode devices**. **a**, Detection voltage ($V_{detect}$) as a function of microwave frequency for various input microwave powers ($P_{RF}$) at zero d.c. bias current. **b**, **c**, $V_{detect}$ as a function of microwave frequency under various d.c. bias currents ($I_{dc}$). The microwave input power ($P_{RF}$) is 0.01 µW. The d.c. bias was found to significantly affect the $V_{detect}$. **d**, Maximum $V_{detect}$ as a function of the d.c. bias current. The inset in Fig. 2d is the detection sensitivity ($V_{detect}/P_{RF}$) as a function of $P_{RF}$ for various d. c. bias currents.

**Figure 3 | A comparison of STT diode sensitivity with the indication of the required bias field**. The STT diode sensitivity values reported previously require an external magnetic field with a particular direction and amplitude, while in this study the giant sensitivity is achieved in the absence of any bias magnetic field, putting the device in the preferred (top-left) corner of the performance graph.

**Figure 4 | Micromagnetic simulations**. **a**, A comparison between precession frequency ($f_p$) and microwave frequency ($f_0$) of maximum detection voltage as a function of $I_{dc}$. **b**, Frequency of the microwave emission in the presence of both microwave and d.c. currents, as a function of the microwave frequency. Inset: Locking region as a function of microwave current amplitude. **c**, Rectified voltage as a function of the microwave frequency ($I_{dc}$ = -0.26 mA and $I_{ac}$ = 0.07 mA). **d**, Intrinsic phase shift between the microwave current and the oscillating TMR signal. This phase introduces a difference between the self-oscillation frequency ($I_{dc}$ = 0 mA) and the frequency of the maximum rectification voltage as shown in Fig. 4a.

**Figure 5 | Microwave emission and static resistance as a function of the d.c. bias current**. **a,** Microwave emission produced by persistent magnetization oscillations. Large output power microwave signals were observed at negative d.c. bias. **b,** Comparison between self-oscillation frequency (olive-color dots) and maximum detection voltage frequency (orange-color dots) as a function of the d.c. bias currents. **c,** Static resistance as a function of d.c. bias. The black (blue) arrows stand for the magnetization vectors of the reference layer (free layer) in MTJ device. In region B the x-component of the magnetization vector of the free layer changes gradually from parallel to antiparallel to the reference layer.

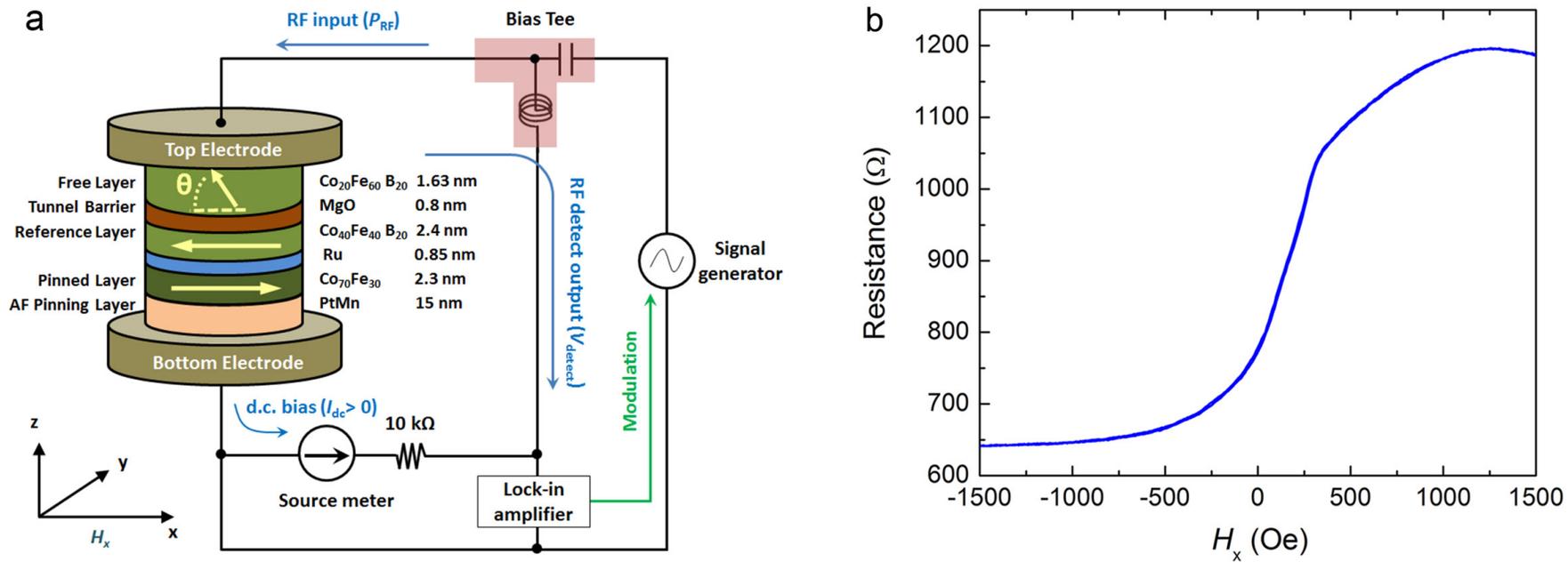

**Figure 1   Bin Fang** *et al*

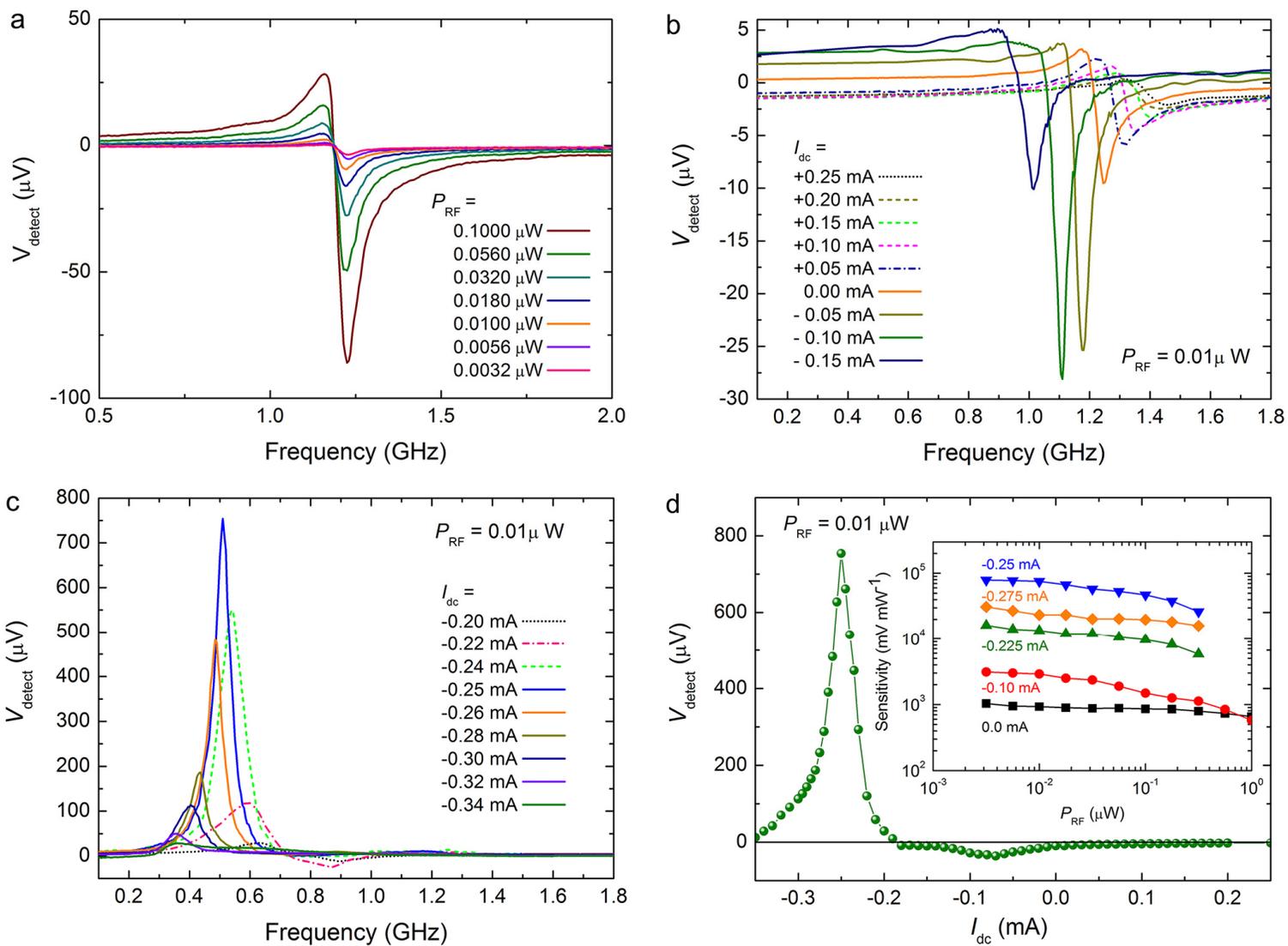

**Figure 2   Bin Fang** *et al*

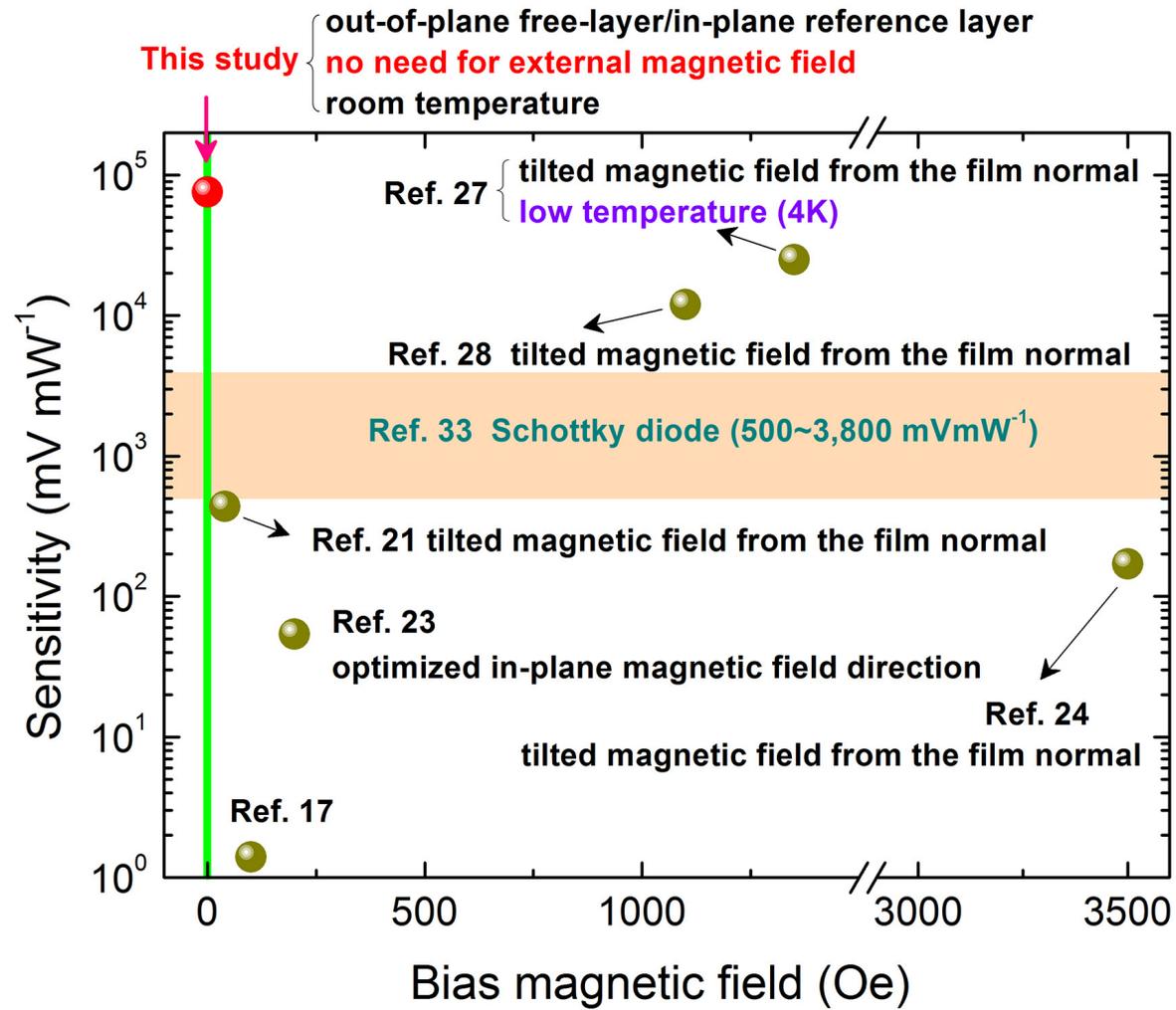

Figure 3   Bin Fang *et al*

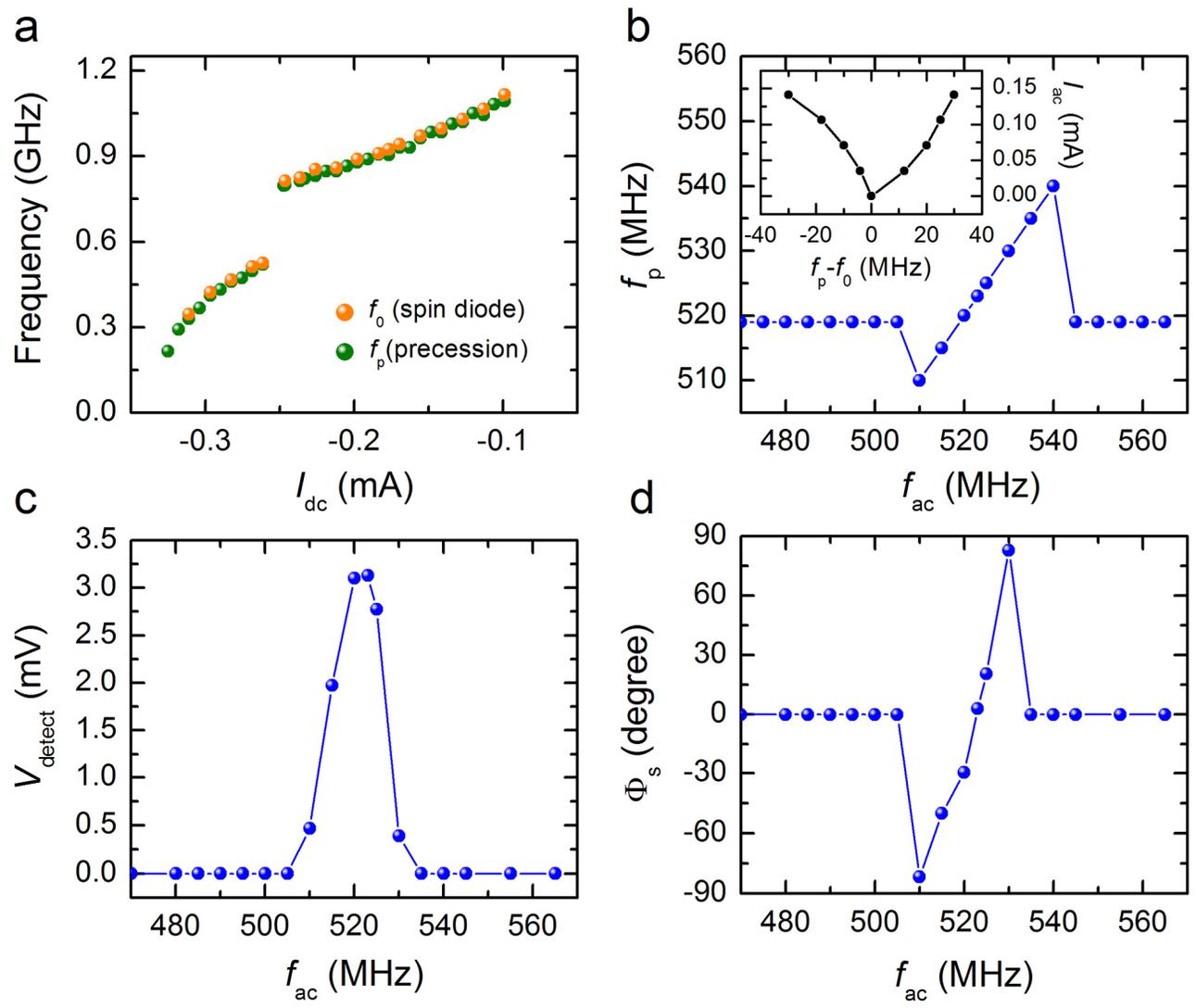

**Figure 4 Bin Fang *et al***

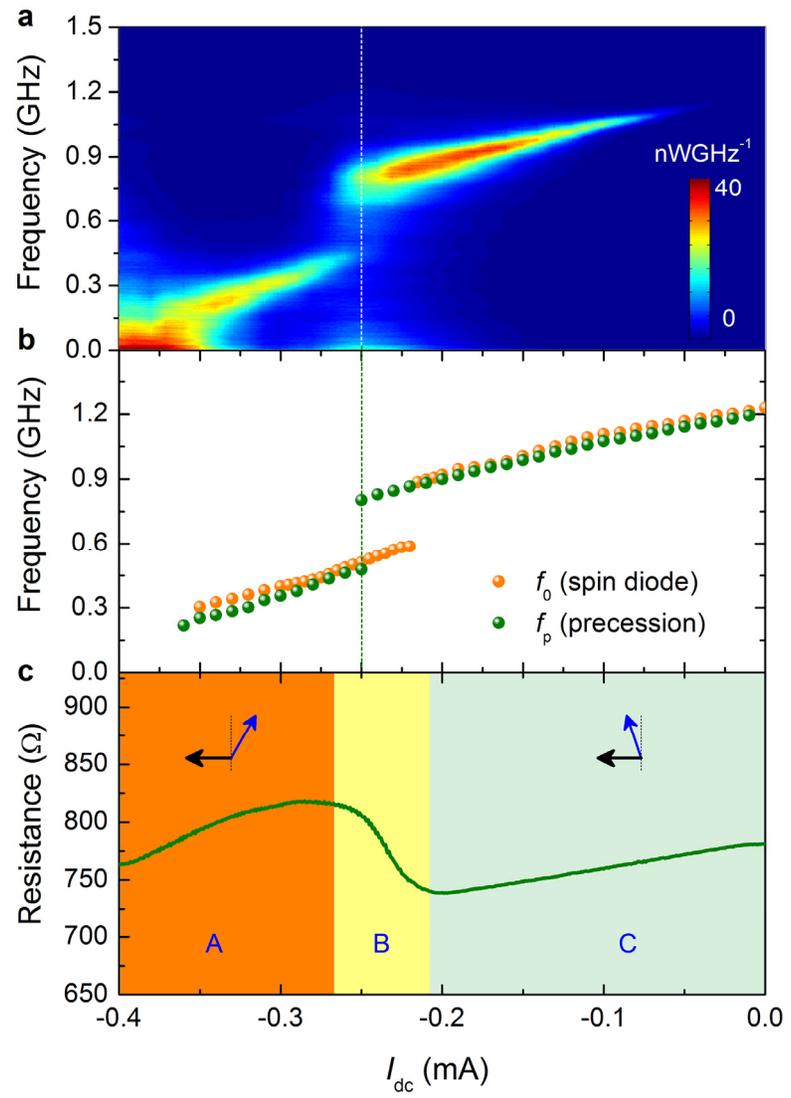

**Figure 5 Bin Fang *et al***

# Supplementary Information

1. **Estimation of the voltage-controlled perpendicular magnetic anisotropy (VCMA)**

The VCMA induced by electric field in our system may increase the sensitivity of the microwave detector. The amplitude of VCMA can be estimated using the method described in Ref. S1. The effective perpendicular magnetic anisotropy energy per unit volume of the free layer, $E_p$, including interface and demagnetization contributions, can be calculated from the area under the in-plane field-magnetoresistance hysteresis loop, whose shape is identical to that of the magnetic tunnel junction (MTJ) conductance $G(H_x)$ hysteresis loop as shown in Fig. S1 (a)

$$E_p = \int_0^{M_s} H_x(M_x)dM_x = \frac{1}{2}M_s H_p, \tag{E1}$$

where $M_s$ is the saturation magnetization of the free layer, $M_x$ is projection of magnetization onto the x axis and $H_p$ is the effective perpendicular anisotropy field. The VCMA induces a change in the slopes of the two $G(H_x)$ curves in Fig. S1(a). Figure S1(b) shows the dependence of the perpendicular magnetic anisotropy field on the applied voltage, where $\Delta H_p(V_{dc}) = H_p(V_{dc}) - H_p(0)$. This dependence is well fit by a straight line with a slope of 0.55 kOe/V, which corresponds to a change in the magnetic anisotropy energy per unit area per applied electric field of 34 fJ/(V · m). The computed experimental parameters are comparable to VCMA values previously observed in similar material systems [S1, S2, S3].

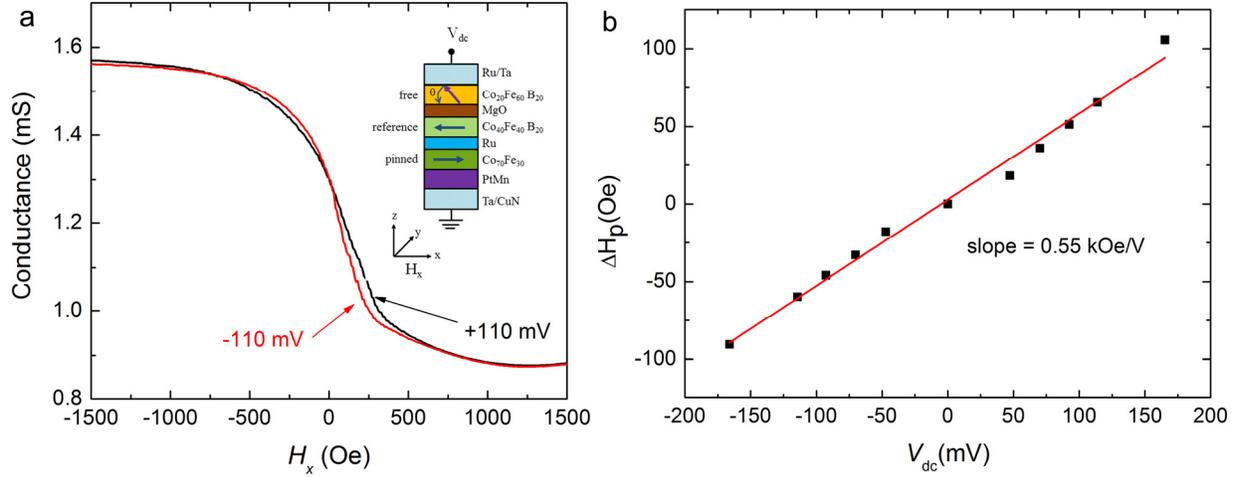

Figure S1. **a,** Hysteresis loop of MTJ conductance (*G*) versus in-plane magnetic field ($H_x$) measured under $\pm 110$ mV direct voltage bias $V_{dc}$. The right inset shows a schematic of the MTJ nanopillar. **b,** Variation of the effective perpendicular anisotropy field $\Delta H_p(V_{dc})$ with d.c. voltage, extracted from the $G(H_x)$ hysteresis loops via Eq. (E1) (squares) and a linear fit to the data (line).

## 2. Measurement setup of spin-transfer-torque microwave emission

We performed STT-induced microwave emission [S4, S5] measurements as shown in Fig. S2 (a). A bias Tee was used to separate the d.c. current injection from the dynamic output signal, which was directly recorded by a 9 kHz-26.5 GHz spectrum analyzer after 30 dB amplification. A baseline taken at $I_{dc}$ = 0 mA was subtracted from the measured data. Figure S2 (b) shows the typical microwave spectra at various d.c. currents. As presented in the main text, a jump in the oscillation frequency is observed, which is related to a change of sign of the *x*-component of the magnetization oscillation axis from positive to negative, with a consequent change in the symmetry of the spin-transfer torque. The reference layer is fixed along the negative *x*-direction, as shown in Fig. 1a in the main text.

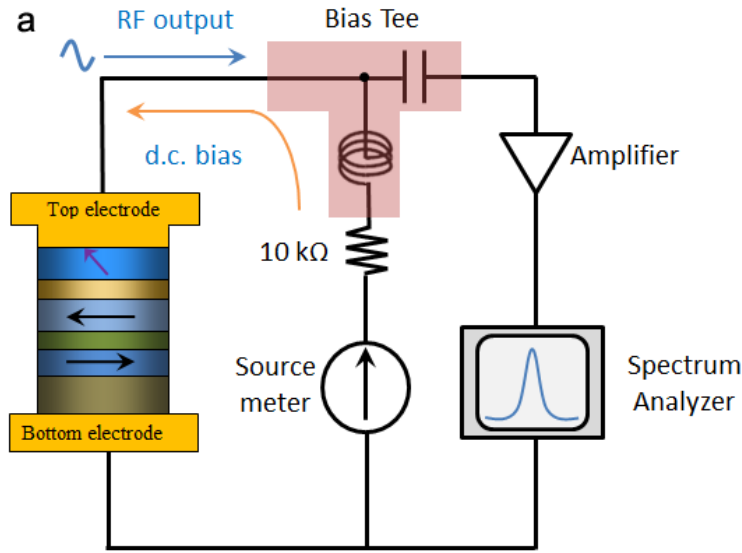

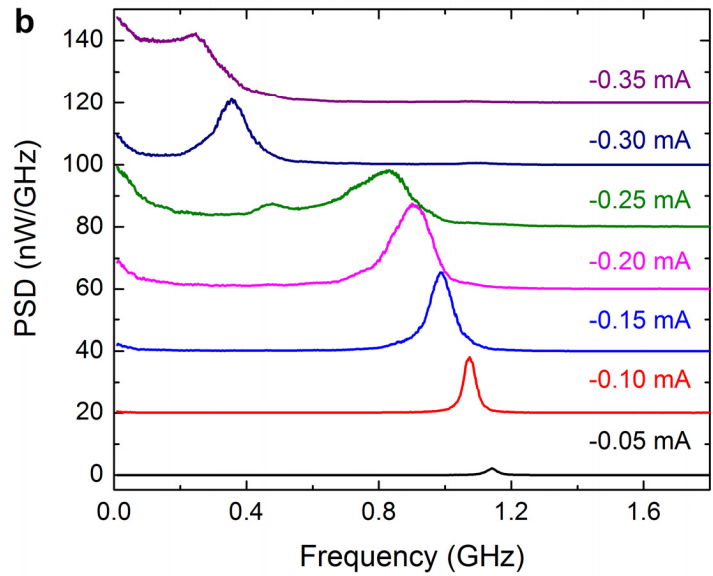

Figure S2. **a**, Schematic of circuit used for STT-induced microwave emission measurements. **b**, Microwave spectra as a function of d.c. bias current $I_{dc}$ at zero applied magnetic field. The curves are offset by approximately 20 nW GHz$^{-1}$ along the vertical axis for clarity.